\documentclass[12pt,preprint]{emulateapj}

\shorttitle{Jupiter's formation in the vicinity of the amorphous ice snowline}
\shortauthors{Mousis et al.}

\usepackage{etex}
\usepackage {amsmath}
\usepackage{color}
\usepackage{array}

\begin{document}


\title{Jupiter's formation in the vicinity of the amorphous ice snowline}


\author{Olivier Mousis\altaffilmark{1}, Thomas Ronnet\altaffilmark{1,2}, and Jonathan I. Lunine\altaffilmark{3}}


\altaffiltext{1}{Aix Marseille Univ, CNRS, CNES, LAM, Marseille, France {\tt olivier.mousis@lam.fr}}
\altaffiltext{2}{Lund Observatory, Department of Astronomy and Theoretical Physics, Lund University, Box 43, 221 00 Lund, Sweden}
\altaffiltext{3}{Department of Astronomy, Cornell University, Ithaca, NY 14853, USA}


\begin{abstract}

Argon, krypton, xenon, carbon, nitrogen, sulfur, and phosphorus have all been measured enriched by a quasi uniform factor in the 2--4 range, compared to their protosolar values, in the atmosphere of Jupiter. To elucidate the origin of these volatile enrichments, we investigate the possibility of inward drift of particles made of amorphous ice and adsorbed volatiles, and their ability to enrich in heavy elements the gas phase of the protosolar nebula once they cross the amorphous-to-crystalline ice transition zone, following the original idea formulated by \cite{Mo15}. To do so, we use a simple accretion disk model coupled to modules depicting the radial evolution of icy particles and vapors, assuming growth, fragmentation and crystallization of amorphous grains. We show that it is possible to accrete supersolar gas from the nebula onto proto-Jupiter's core to form its envelope, and allowing it to match the observed volatile enrichments. Our calculations suggest that nebular gas with a metallicity similar to that measured in Jupiter can be accreted by its envelope if the planet formed in the $\sim$0.5--2 Myr time range and in the 0.5--20 AU distance range from the Sun, depending on the adopted viscosity parameter of the disk. These values match a wide range of Jupiter's formation scenarios, including in situ formation and migration/formation models. 

\end{abstract}


\section{Introduction}

\begin{figure}[h]
\begin{center}
\resizebox{\hsize}{!}{\includegraphics[angle=0]{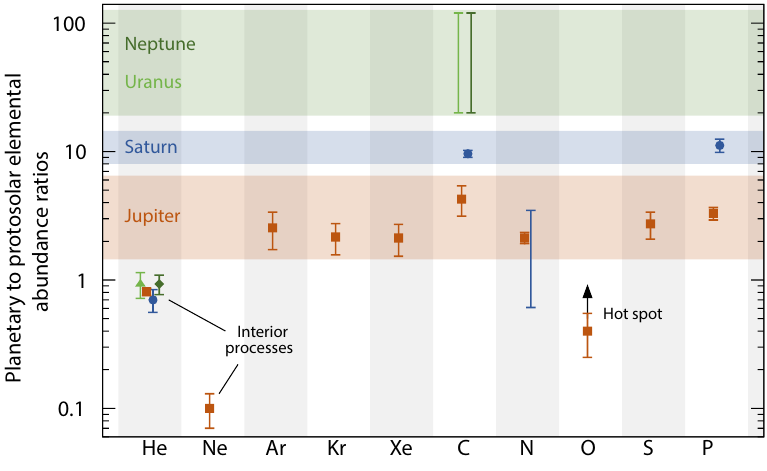}}
\caption{Enrichment factors (with respect to the protosolar value) of noble gases and heavy elements measured in Jupiter, Saturn, Uranus, and Neptune. Error bars, central values and planets share the same color codes. The helium determination is taken from the Galileo probe in situ measurements at Jupiter \citep{vo98,Ni98}, from Voyager measurements at Saturn and Uranus \citep{Co00,Co87,At18}, and from ISO measurements at Neptune \citep{Bu03}. The neon, argon, krypton, xenon and carbon determinations at Jupiter correspond to Galileo probe in situ measurements \citep{Ma00,Wo04}. The carbon determination is taken from Cassini infrared measurements at Saturn \citep{Fl09a}, and from Voyager 2 and HST measurements at Uranus and Neptune \citep{Li87,Li90,ba95,Ka09,Ka11,Sr14}. The nitrogen determination is derived from Juno measurements at pressures of 100 bars or more at Jupiter \citep{Bo17}, and from Cassini measurements at Saturn \citep{Fl11}. The oxygen and sulfur determinations are based on Galileo probe in situ measurements at Jupiter \citep{Wo04} (probably a lower limit for oxygen, not representative of the bulk O/H). The phosphorus determinations at Jupiter and Saturn are taken from Cassini measurements \citep{Fl09b}. We refer the reader to \cite{Mo18} for further details about the calculations of the error bars.}
\label{fig1}
\end{center}
\end{figure}

The source of the volatile enrichments measured in the atmospheres of the four giant planets of the solar system is still matter of debate. Due to the absence of in situ explorations of Saturn, Uranus and Neptune, the number of data regarding their compositions remains scarce, but suggests significant C enrichments in their atmospheres, relatives to the protosolar value (see Fig. \ref{fig1}). In the case of Jupiter, thanks to the mass spectrometry data returned by the {\it Galileo} probe down to the $\sim$20 bar region, Ar, Kr, Xe, C, N, and S have all been found enriched by a quasi uniform factor in the 2--4 range, compared to their protosolar values \citep{Ni98,Ma00,Wo04}. Infrared spectroscopy observations by the {\it Galileo} orbiter and later by the {\it Cassini} spacecraft also allowed to retrieve a P abundance enriched by a similar factor in Jupiter's atmosphere \citep{Ir98,Fl09b}. The {\it Juno} mission recently provided a new deep N mixing ratio, which is still enriched relative to the protosolar value, but slightly lower than the one previously found by the {\it Galileo} probe (350 $\pm$ 20 ppmv vs. 664 $\pm$ 254 ppmv) \citep{Bo17,Wo04}. Contrasting with these measurements, the {\it Galileo} probe also found that the He, Ne and O abundances are depleted in Jupiter's atmosphere, compared to their protosolar values. He and Ne depletions have been attributed to precipitation of He droplets in the deep atmosphere \citep{St77a,St77b} and Ne sequestration within these droplets \citep{Wi10}. The O depletion is often attributed to the dynamics of the region within which the Galileo probe descended \citep{Or98} or might reflect the bulk composition of the planet \citep{Mo12}.

To explain Jupiter's properties, it has been proposed that its atmosphere reflects the composition of cold planetesimals made of amorphous ice accreted during the growth of the planet \citep{Ow99}. Alternative scenarios suggest that the metallicity of the planet's envelope results from the accretion of planetesimals made of clathrates and/or crystalline ices \citep{Ga01,He04,Ga05,Mo09,Mo12,Mo14}. Another category of proposed models are those advocating that the giant planet's envelope was fed from protosolar nebula (PSN) gas already enriched in heavy elements. For instance, the model proposed by \cite{Gu06} suggests that far ultraviolet (FUV) photoevaporation would have depleted the disk in He, He and Ne in regions where all the other species have condensed into icy grains, precluding them from being swept up in the photoevaporative flow. Inward drift of icy grains crystallized at low temperature and their subsequent vaporization in hotter disk regions would have then favored the delivery of a gas enriched in Ar, Kr, and Xe to the envelope of the forming Jupiter. \cite{Mo15} subsequently improved the model of \cite{Gu06} by focusing on the trapping conditions of the volatiles in the outer regions of the PSN. They find that the FUV flux generating photoevaporation in the outer PSN also enables an efficient trapping of the volatiles in amorphous water. Because volatiles trapped together by amorphous water are likely to be released together at a similar temperature, \cite{Mo15} point out that the trapped volatiles would have the same evaporation radii in the PSN, thus potentially leading to  enrichments in volatiles in the disk gas phase that would be similar to those observed in Jupiter (see Fig. \ref{fig1}). In this scenario, Jupiter's current metallicity would have essentially been acquired during its accretion from PSN gas. 

In addition to the idea formulated by \cite{Mo15}, two other mechanisms can potentially provide supersolar PSN gas to the forming Jupiter. The inward migration of crystalline ices through their different snowlines can lead to the formation of metal-rich gases in the outer PSN, as suggested by the results obtained by \cite{Boo17}, based on the consideration of the evolutions of H$_2$O, CO, and CO$_2$ vapors close to their respective ice lines. Also, even if it remains to be demonstrated, metals could be added to the PSN gas by the destabilization of clathrate hydrates formed from solid crystalline ice and vapors in the outer PSN.

\begin{figure}
\begin{center}
\resizebox{\hsize}{!}{\includegraphics[angle=0]{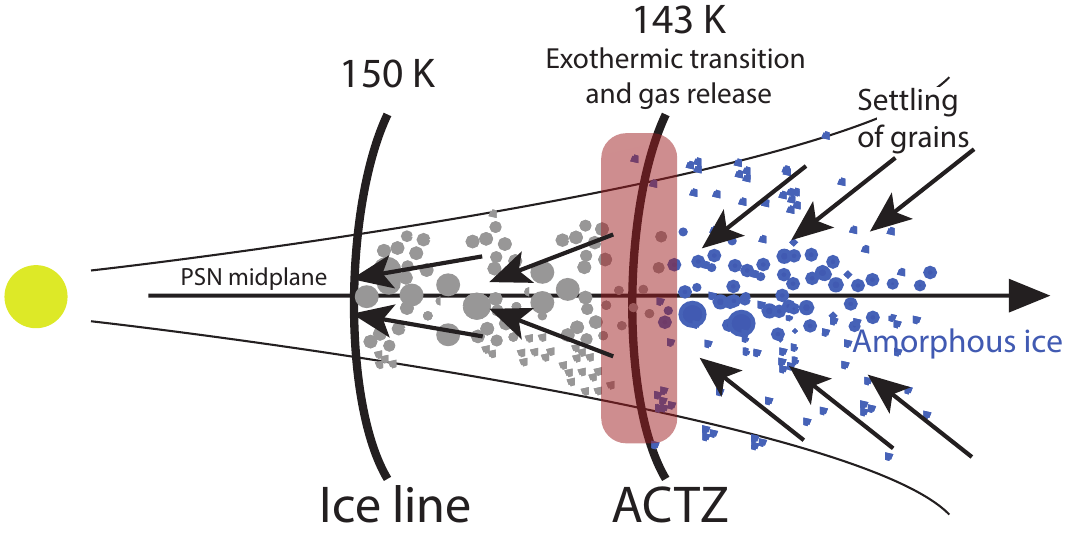}}
\caption{Sketch illustrating the scenario to explain a homogeneous enrichment in volatiles in the envelope of Jupiter. A gaseous protosolar disk is shown edge-on. Black arrows represent the dynamical evolution of grains (sedimentation, coagulation, and inward drift). Pristine amorphous particles (blue color) drift inward from the outer PSN and cross the ACTZ, which corresponds to a transition temperature of $\sim$143 K \citep{Ko90,Ba07}. Once crystallized, these particles (grey color) continue their inward drift and  release the adsorbed volatiles to the PSN gas phase. The metallicity of the PSN gas phase progressively increases over several AU around the ACTZ and may reach values comparable with the one measured in Jupiter.}
\label{fig2}
\end{center}
\end{figure}

In the present work, we consider the idea of \cite{Mo15} and investigate the possibility of inward drift of particles made of amorphous ice and adsorbed volatiles and their ability to enrich in Ar, Kr, Xe, C, N, S, and P the gas phase of the PSN located around the snowline corresponding to the amorphous-to-crystalline ice transition zone (ACTZ) (see Fig. \ref{fig2}). This transition enables the direct release of adsorbed volatiles in the PSN gas phase at $\sim$143 K \citep{Ko90,Ba07}. To do so, we use a simple accretion disk model coupled to modules that track the radial evolutions of icy particles and vapors, assuming growth, fragmentation and crystallization of amorphous grains. We don't make any assumption about the origin of these particles since they can either originate from the ISM in which the mantles of icy grains are dominated by amorphous ice \citep{Gi04} or been generated by UV photons in the outer PSN \citep{Ci14}. 

\section{Model}

\subsection{Accretion disk}

We model the evolution of the PSN through the decrease of the mass accretion rate $\dot{M}$ of the disk over time as constrained by observations \citep{Ha98} 

\begin{equation}
\label{Mdot}
\log\left( \frac{\dot{M}}{M_\odot/\mathrm{yr}}\right) = -8.00-1.40\log\left(\frac{t+0.1}{\mathrm{Myr}}\right),
\end{equation}

\noindent where $M_\odot$ is the mass of the Sun. The mass accretion rate can be related to the viscosity $\nu$ and surface density $\Sigma_g$ of the disk through

\begin{equation}
\label{visc}
\dot{M} = 3\pi\nu\Sigma_g,
\end{equation}

\noindent where $\dot{M}$ is assumed to be constant with the heliocentric distance. The viscosity of the disk is expressed as $\nu = \alpha C_s^2/\Omega_K$ \citep{Sh73}. Here, $\Omega_K$~=~$\sqrt{G M_\odot / r^3}$ is the keplerian frequency with $G$ defined as the gravitational constant, and $C_s$ is the isothermal sound speed given by

\begin{equation}
\label{sound}
c_s = \sqrt{\frac{k T}{\mu m_p}},
\end{equation}

\noindent where $k$ is the Boltzmann constant, $\mu$ is the mean molecular weight (for our purposes $\mu$ = 2.3) and $m_p$ is the proton mass. The value of $\alpha$ is a free parameter measuring the turbulence strength which is regulating the efficiency of viscous heating and therefore the temperature of the disk.

\begin{figure}
\begin{center}
\resizebox{\hsize}{!}{\includegraphics[angle=0]{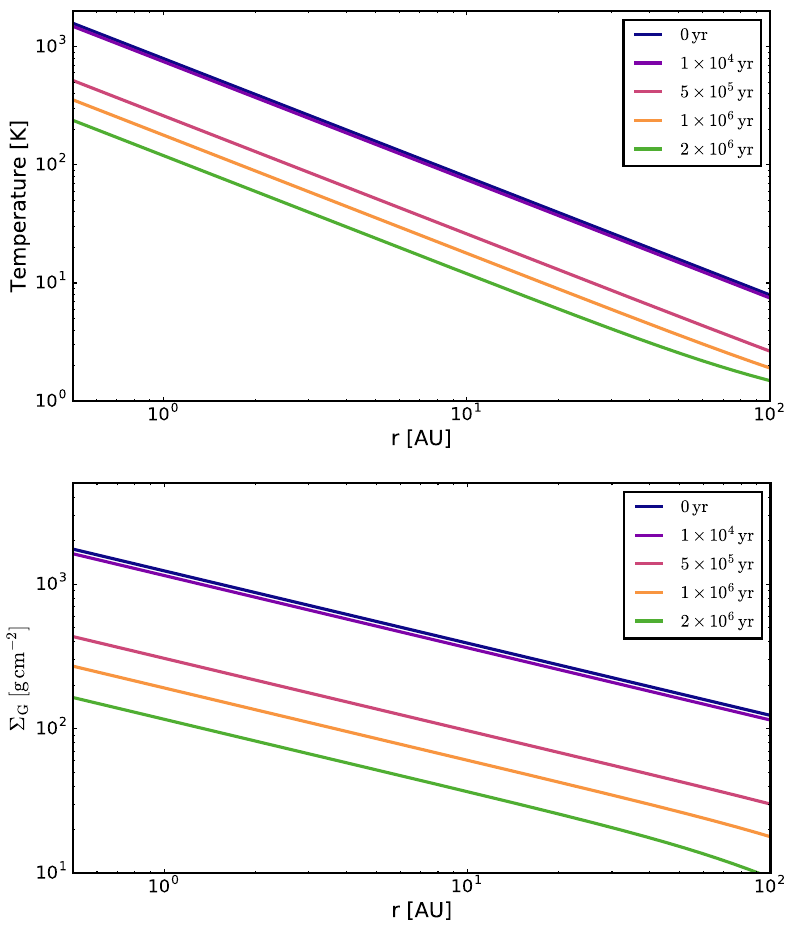}}
\caption{PSN temperature and surface density profiles calculated at $t$ = 0, 10$^4$, 5~$\times$~10$^5$, 10$^6$, and 2~$\times$~10$^6$ yr for $\alpha$~=~0.01.}
\label{fig3}
\end{center}
\end{figure}

The temperature $T_d$ at the midplane of the disk is computed from the analytical expression derived by \cite{NN94}:

\begin{equation}
\label{temp}
\sigma_\mathrm{sb} T^4_d = \frac{1}{2}\left(\frac{3}{8} \tau_\mathrm{R}+\frac{1}{2\tau_\mathrm{P}}\right)\Sigma_g\nu\Omega^2_K,
\end{equation}

\noindent where $\sigma_\mathrm{sb}$ is Stefan-Boltzmann constant, and $\tau_\mathrm{R}$ and $\tau_\mathrm{P}$ are the Rosseland and Planck mean optical depths, respectively. We assume $\tau_\mathrm{P}=2.4\tau_\mathrm{R}$ \citep{NN94}. $\tau_\mathrm{R}$ is derived from \citep{Hu05}:

\begin{equation}
\tau_\mathrm{R}=\frac{\kappa_\mathrm{R}\Sigma_g}{2},
\end{equation}

\noindent where $\kappa_\mathrm{R}$ is the the Rosseland mean opacity. We use a simple expression for the opacity appropriate for silicate dust, namely $\kappa_\mathrm{R}=\kappa_0T^{1/2}_d$ with $\kappa_0 = 1\times10^{-2}$ m$^2 \,$kg$^{-1}$ \citep{Ki12}. Equations~(\ref{visc}) and (\ref{temp}) are solved together iteratively before the disk is evolved in time via Eq.~(\ref{Mdot}), and the processus is repeated. Figure \ref{fig3} shows the PSN temperature and surface density profiles at different epochs for $\alpha$ = 0.01.

\subsection{Evolution of dust size}

Our dust evolution model is based on the works of \cite{Bi12} and \cite{La14}. We assume that dust is initially present in the disk in the form of micron-sized amorphous icy grains that grow through mutual collisions on a timescale given by:

\begin{equation}
\tau_\mathrm{growth} =\frac{a}{\dot{a}}= \frac{4\Sigma_d}{\sqrt{3}\epsilon_g\Sigma_g\Omega_K},
\end{equation}

\noindent where $a$ is the size of the dust grains, $\Sigma_d$ is the dust surface density assuming $\Sigma_d/\Sigma_g$ = 0.01 at $t$ = 0, and $\epsilon_g=0.05$ is a parameter accounting for the growth/sticking efficiency \citep{La14}. The maximum size that grains are able to reach is limited by several growth barriers. Here, following \cite{Bi12}, we only account for the size limitations of dust grains due to fragmentation and radial drift. Fragmentation occurs when the relative velocity of dust grains, due to turbulent motion, exceeds the fragmentation velocity threshold $u_\mathrm{f}$. In this case, the representative dust grain size is \citep{Bi12}: 

\begin{equation}
 a_\mathrm{frag}=f_\mathrm{f}\frac{2\Sigma_g}{3\pi\rho_s\alpha}\frac{u^2_\mathrm{f}}{c^2_s},
\end{equation}   

\noindent where the prefactor $f_\mathrm{f}=0.37$ accounts for the fact that the representative size of the dust grains is slightly below the maximum reachable size, $\rho_s = 1\, \mathrm{g\,cm}^{-3}$ is the dust internal density, and $u_\mathrm{f}$ is set to $10\,\mathrm{m\,s}^{-1}$.

In the case where the size of the grains is limited by their drift (i.e., the grains drift inward before being able to grow further) their representative size is

\begin{equation}
 a_\mathrm{drift}=f_\mathrm{d}\frac{2\Sigma_d}{\pi\rho_s}\frac{v^2_K}{c^2_s} \left|\frac{\mathrm{d}\ln P}{\mathrm{d}\ln r}\right|^{-1},
\end{equation}

\noindent where again the prefactor $f_\mathrm{d}=0.55$ accounts for the shift of the representative size as compared to the maximum size of the dust grains \citep{Bi12}, $v_K$ is the keplerian velocity, $P = c^2_s\rho_g$ is the gas pressure at the midplane of the disk and $\rho_g = \Sigma_g/\sqrt{2\pi}H_g$ the gas density in the the disk's midplane. These two limiting sizes are calculated at each orbital distance and the minimum value sets the size up to which the grains are allowed to grow. 

\subsection{Evolution of vapor and dust}

Here, we assume that Ar, Kr, Xe, C, N, S, and P are simultaneously released as vapors during the crystallization of amorphous icy grains when they cross the ACTZ at a temperature of $\sim$143 K in the PSN \citep{Ko90,Ba07}. We also assume that the gaseous mixture released by the crystallizing amorphous ice has a protosolar composition. Once released, these vapors form a homogeneous gaseous mixture that radially diffuses and advects. The surface density $\Sigma_i$ of a trace species $i$ (e.g., vapor, small or large dust grains) is evolved by numerically solving the following 1D radial advection-diffusion equation \citep{Bi12}:

\begin{equation}
\frac{\partial\Sigma_i}{\partial t} + \frac{1}{r}\frac{\partial}{\partial r}\left[r\left(\Sigma_i v_i - D_i \Sigma_g \frac{\partial}{\partial r}\left(\frac{\Sigma_i}{\Sigma_g}\right)\right)\right] + \dot{Q} = 0,
\end{equation}

\noindent where $v_i$ and $D_i$ are the radial velocities and diffusivities of species $i$, respectively. $\dot{Q}$ corresponds i) to the source term of species $i$ vapor released to gas and ii) to the sink term of species $i$ in solid phase when the icy grains have drifted through the ACTZ. It is given by
\begin{equation}
  \dot{Q}_\mathrm{vap}=-\dot{Q}_\mathrm{ice} = \left \lbrace \begin{array}{ll}
  \frac{\Sigma_\mathrm{ice}}{\Delta t} & \mathrm{for} \; T_d \geqslant 143\,\mathrm{K}\\
  {} & {} \\
  0 & \mathrm{for} \; T_d < 143\,\mathrm{K}, \\ 
  \end{array} \right.
\end{equation} 
\noindent where $\Delta t$ is the timestep of the simulation.

The velocity of the dust is \citep{Bi12}:

\begin{equation}
v_\mathrm{d} = -\frac{2\mathrm{St}}{1+\mathrm{St}^2}\eta v_K + \frac{1}{1+\mathrm{St}^2}v_g,
\end{equation}

\noindent where St is the Stokes number of the particles describing their aerodynamic properties, which, assuming an Epstein drag law, is

\begin{equation}
\mathrm{St}=\frac{a\pi\rho_s}{2\Sigma_g}.
\end{equation}
The radial (inward) velocity of the gas in a quasi-stationary disk is given by 

\begin{equation}
\label{Vgas}
v_g = -\frac{3}{2}\frac{\nu}{r}.
\end{equation}

\noindent We used this velocity for vapor species. The diffusivity of vapor species is assumed to be that of the gas $D_g = \nu$. The diffusivity of dust is in turn given by:

\begin{equation}
D_\mathrm{d} = \frac{D_g}{1+\mathrm{St}^2}.
\end{equation}

\section{Results}

\begin{figure}
\begin{center}
\resizebox{\hsize}{!}{\includegraphics[angle=0]{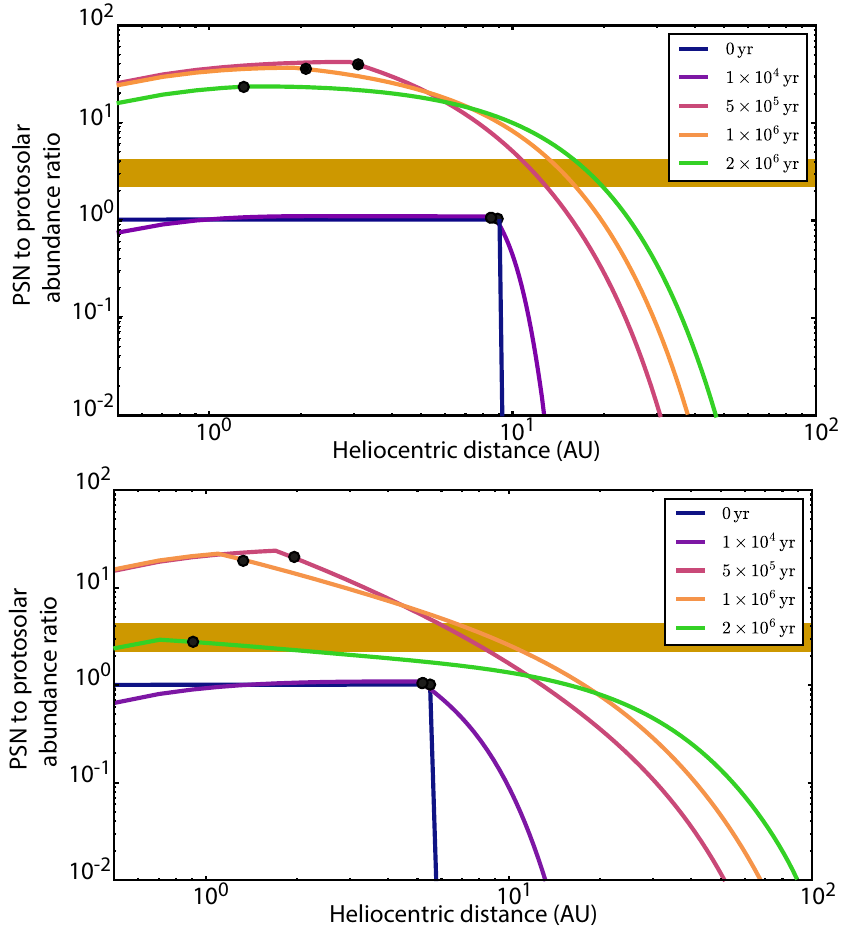}}
\caption{Time and radial evolution of the abundances of volatiles released to the PSN gas phase by the icy grains subsequent to their drift through the ACTZ. Calculations have been performed for $\alpha$ = 10$^{-3}$ (top panel) and $\alpha$ = 10$^{-2}$ (bottom panel). The brown horizontal bar represents the range of volatile enrichments (nominal values) measured in Jupiter (see Fig. \ref{fig1}). At a given epoch, the metallicity of the PSN gas phase matches Jupiter's value at heliocentric distances at which the abundance profile of volatiles intercepts the horizontal bar. The black dots designate the location of the ACTZ ($\sim$143 K) during the evolution of the PSN.}
\label{fig4}
\end{center}
\end{figure}

Figure \ref{fig4} represents the evolution of the volatile abundances normalized to protosolar in the PSN and supplied as vapors to the gas phase for two different values of the viscosity parameter $\alpha$, namely 10$^{-3}$ and 10$^{-2}$. Volatiles are released to the gas phase of the PSN by the drifting grains once they cross the ACTZ and diffuse radially. With time, the ACTZ moves inward and evolves from $\sim$9 AU ($\alpha = 10^{-3}$) and $\sim$5.5 AU ($\alpha = 10^{-2}$) at $t$~=~0 down to close 1 AU after 2 Myr of PSN evolution. Because of the initially high surface density of dust in the PSN (1\% that of the gas), its gas phase around the location of the ACTZ is quickly enriched in volatiles released by the crystallization of drifting amorphous particles. The volatile enrichments factors (relatives to their protosolar values) reach maximums of $\sim$40 and 25 at $t$~=~0.5 Myr in the $\sim$2-3 AU region of the PSN, which corresponds to the location of the ACTZ at this epoch, assuming $\alpha$ values of 10$^{-3}$ and 10$^{-2}$, respectively. The enrichments profiles decrease with time because of the diminishing number of particles crossing the ACTZ and releasing their volatiles. This trend is faster when using a higher $\alpha$ value, as diffusion is more efficient.

Figure \ref{fig5} shows the evolution of the volatile abundances normalized to protosolar in the PSN and remaining trapped in the icy particles transported throughout the disk via diffusion and gas drag until they reach the ACTZ. At early epochs, the dust surface density is large, implying an efficient growth of the grains up to the pebbles size in the outer regions of the PSN, and consequently a fast drift of these particles inward the disk as they reach larger sizes (up to a few centimeters). With time, the dust surface density decreases in the PSN and so does the dust mass flux through the disk. Interestingly, while less prominent than in the previous case, the volatile abundances supplied by solids in the PSN can be enhanced by factors up to $\sim$3 and 14, compared to their protosolar values at $t$~=~0.5 Myr in the $\sim$3--4 AU region of the PSN, for $\alpha$ values of 10$^{-3}$ and 10$^{-2}$, respectively. 

Our calculations suggest that it is possible to accrete PSN gas with a metallicity similar to that measured in Jupiter if its envelope forms between $\sim$0.5 and 2 Myr between 0.5 and 20 AU from the Sun, depending on the adopted $\alpha$ value. These values match a rather wide range of Jupiter's formation scenarios, including in situ formation at 5 AU \citep{Po96} and migration/formation models \citep{Al05}.


\section{Conclusions}

\cite{Mo15} initially proposed that amorphous ice migrating inward the PSN could lose its volatile content and then enrich the gas phase of Jupiter’s feeding zone. Here, we built a numerical model allowing us to quantitatively test this hypothesis and we find that it is indeed possible to accrete supersolar gas from the PSN onto proto-Jupiter's core to form its envelope, and allowing it to match the observed volatile enrichments. This implies that there is no need to invoke the accretion of solids \citep{Ow99,Ga01,Ga05,Mo09,Mo12,Mo14} or the core erosion to provide the required amounts of Ar, Kr, Xe, C, N, S, and P in Jupiter's envelope \citep{Wi10}. The delivery of heavy elements trapped in icy solids to proto-Jupiter's envelope can be a concern since the accretion of pebbles to form the core creates a gap in the pebble disk and halts the accretion of pebbles prior the collapse of the gas envelope \citep{Lam14}. A postformation accretion of planetesimals by Jupiter would be also extremely limited given their high probabilities of ejection vs. accretion \citep{Gu00}. Also, based on dynamical simulations, \cite{Ro18} find that, to reproduce both masses of the Galilean satellites and the main asteroid belt via the implantation of planetesimals originating from the outer edge of Jupiter's gap, no much more than a mass equivalent to that of the Galilean system (a few 10$^{-4}$ Jupiter masses) can be captured by Jupiter. We thus find that the formation of Jupiter's envelope from supersolar PSN gas resulting from the release of vapors due to the crystallization of amorphous ice avoids the problem of solids accretion. 
 
 Our formalism presupposes that the gas accretion rate is inward at all radii (see Eq. \ref{Vgas}). However, viscous disks theory predicts that the gas flow becomes outward at a given transition radius. This effect may have an impact on the evolution of the gas metallicity of the disk, in particular if the transition radius becomes close to the ACTZ in the PSN. We have investigated the influence of this effect on our model by adopting the expression of gas radial velocity derived by \cite{Ly74}, see also \cite{Mor16}, for the evolution of solids and gas:
 
\begin{equation}
 v_g=-\frac{3}{2} \frac{\nu}{r} \left[1-\frac{4(G M_\odot r)^2}{\tau} \right]
 \end{equation}
 
 \noindent where  $\tau$ is a normalized time, defined as
 
\begin{equation}
\tau = 12(GM_\odot)^2 \nu t + 1.
\end{equation}

\noindent Based on this formalism, we find that the transition radius between the inward and outward gas flows always remains far from the ACTZ location. For example, assuming $\alpha$ = 0.01, the ACTZ location decreases from $\sim$5.5 AU to 0.9 AU while the transition radius expands from 15.1 AU to 32.9 AU over the first 2 Myr of the PSN evolution. As a result, we find that the presence of this transition radius has little effect on our results, which still display significant enrichments in volatiles around the ACTZ location in the PSN.

It has been suggested that photoevaporation could enhance the formation of amorphous ice at the outer edge of the PSN \citep{Mo15}. While photoevaporation can potentially increase the disk's metallicity \citep{Gu06,Mo15}, we point out that the formation of amorphous ice via this mechanism is not a requirement. This form of ice is ubiquitous in the ISM \citep{Gi04} and should be also prominent in the presolar cloud, as well as at early epochs in the outer regions of the PSN. Rosetta observations of comet 67P/Churyumov-Gerasimenko's composition also support the idea that early ice was amorphous in the PSN \citep{Mo16}. Depending on the regions toward which the amorphous ice grains drifted, they may have undergone crystallization at the ACTZ location, allowing clathrate hydrate formation during the disk cooling. Comet 67P/Churyumov-Gerasimenko is presumably agglomerated from clathrate hydrates formed under these peculiar circumstances \citep{Mo16}, but other comets formed at further distances from the Sun might have agglomerated from amorphous ice.

\begin{figure}
\begin{center}
\resizebox{\hsize}{!}{\includegraphics[angle=0]{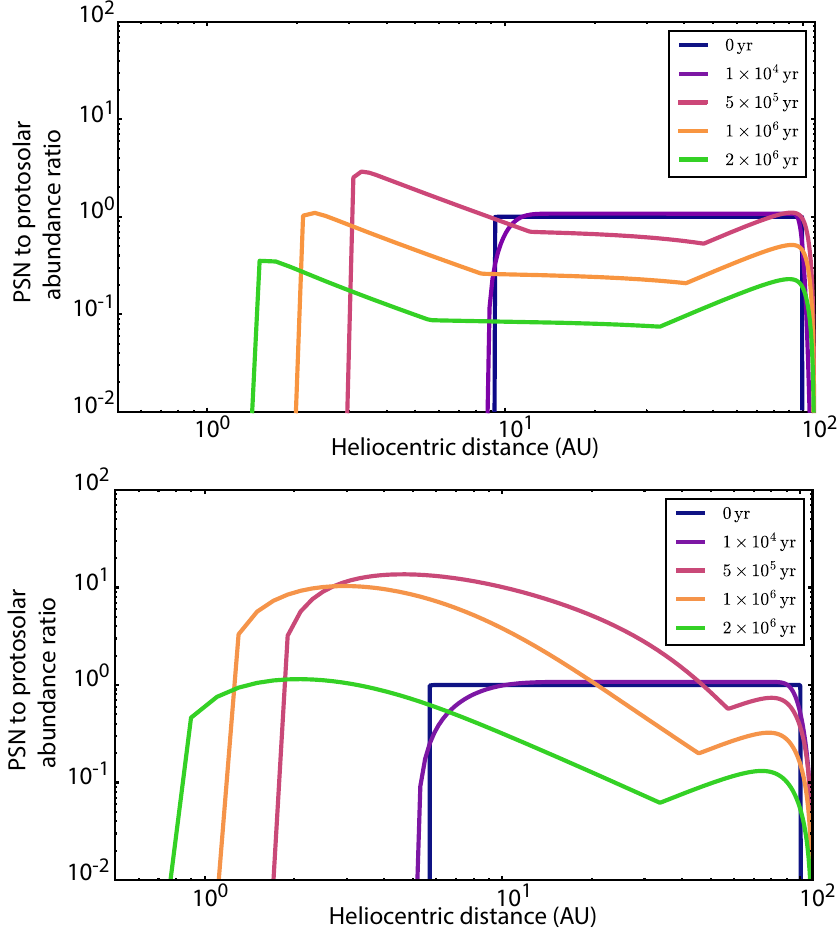}}
\caption{Time and radial evolution of the abundances of volatiles remaining in icy grains prior to their drift through the ACTZ. Calculations have been performed for $\alpha$ = 10$^{-3}$ (top panel) and $\alpha$~=~10$^{-2}$ (bottom panel). With time, the transport of solid grains can also generate supersolar metallicities over several AU in the PSN. However, the delivery of solids including volatiles to proto-Jupiter's envelope may be halted because of the creation of a gap around the planet (see text).}
\label{fig5}
\end{center}
\end{figure}

Interestingly, the contribution of photoevaporation to Jupiter and Saturn's supersolar metallicities via the delivery of gas with an enhanced dust-to-gas ratio to their envelopes, needs to be quantified because gap formation halts the accretion of pebbles \citep{La14}. Since both Jupiter and Saturn have probably created gaps during their formation \citep{At18}, their supersolar metallicities could only be explained either i) via core erosion or ii) via the delivery of supersolar gas to the envelope, as investigated in this work. In the latter case, photoevaporation could help increasing the disk’s gas metallicity thanks to the inward drift of a larger concentration of pebbles throughout the ACTZ. Another potential issue is the competition between the inward drift of pebbles/particles with the increase of the dust-to-gas ratio in the outer disk, due to photoevaporation. Figure \ref{fig5} shows that the abundance of volatiles remaining in solid form becomes significantly reduced by factors reaching $\sim$10 after 2 Myr of the PSN evolution in the $\sim$10--100 AU region. If giant planets formed in these regions of the PSN and acquired their supersolar metallicities through the accretion of gas and gas-coupled particles, then photoevaporation has to be very efficient to counterbalance the cleaning of the outer disk in solid particles.

\begin{figure}[h]
\begin{center}
\resizebox{\hsize}{!}{\includegraphics[angle=0]{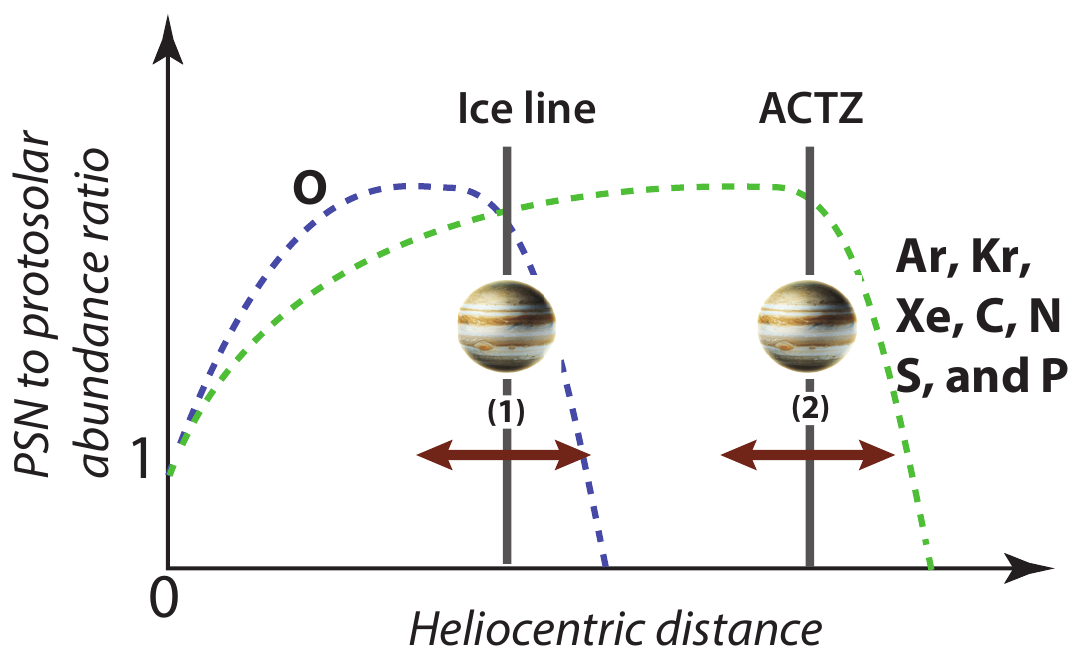}}
\caption{Influence of Jupiter's formation location on the oxygen content in its envelope, assuming that H$_2$O is the main oxygen--bearing volatile in the PSN (see text). Here, Jupiter's feeding zone contains water in both solid and vapor forms while the other volatiles remain exclusively in vapor phase once released from the amorphous particles crossing the ACTZ. Two extreme cases can be envisaged for the oxygen abundance in Jupiter's envelope: (1) Jupiter's formation around the ice line where the oxygen abundance is supersolar, and (2) formation around the ACTZ where Jupiter's oxygen abundance is smaller, and eventually subsolar (see text for details).}
\label{fig6}
\end{center}
\end{figure}

While the Ar, Kr, Xe, C, N, S, and P enrichments in Jupiter can be explained via its formation over a large range of heliocentric distances in the PSN, it appears that the O abundance in the envelope strongly depends on its formation location. Figure \ref{fig6} is a schematic diagram representing the influence of Jupiter's formation location on the O content in its envelope. From this figure, two extreme scenarios of O abundance can be envisaged in Jupiter's envelope: (1) formation around the ice line where Jupiter's oxygen abundance is supersolar due to the redistributive diffusion of water vapor around its vaporization location, and (2) formation around the ACTZ where Jupiter's oxygen abundance is smaller, and eventually subsolar, because of the limited amount of extra water supplied by the outward diffusion of vapor. Here, the oxygen abundance depends on the condensation/coagulation rate of H$_2$O particles beyond the ice line and their ability to be accreted by the envelope of proto-Jupiter \citep{Lam14}, and the amount of O locked in gaseous CO that is enriched relative to protosolar CO at the ACTZ location. Both cases match the Ar, Kr, Xe, C, N, and P enrichments observed in Jupiter's atmosphere. Dedicated simulations, including the development of a module depicting i) the rates of condensation/coagulation of particles around the iceline, ii) their ability to be accreted in the envelope of proto-Jupiter, and iii) disk photoevaporation, will have to be performed after the determination of the deep O abundance in Jupiter by the NASA Juno spacecraft to disentangle these different scenarios.

\acknowledgements
O.M. acknowledges support from CNES. O.M. and T.R. acknowledge support from the A*MIDEX project (n\textsuperscript{o} ANR-11-IDEX-0001-02) funded by the ``Investissements d'Avenir'' French Government program, managed by the French National Research Agency (ANR). J.I.L thanks the Juno project for support.



\newpage 

\newpage 

\newpage 

\newpage 

\newpage 

\newpage 

\begin{thebibliography}{}

\bibitem[Alibert et al.(2005)]{Al05} Alibert, Y., Mousis, O., Mordasini, C., \& Benz, W.\ 2005, \apjl, 626, L57 

\bibitem[Atreya et al.(2018)]{At18} Atreya, S.~K., Crida, A., Guillot, T., et al.\ 2018, In Saturn in the 21st century, Cambridge University Press, in press.

\bibitem[Bar-Nun et al.(2007)]{Ba07} Bar-Nun, A., Notesco, G., \& Owen, T.\ 2007, \icarus, 190, 655 

\bibitem[Baines et al.(1995)]{ba95} Baines, K.~H., Mickelson, M.~E., Larson, L.~E., \& Ferguson, D.~W.\ 1995, \icarus, 114, 328 

\bibitem[Birnstiel et al.(2012)]{Bi12} Birnstiel, T., Klahr, H., \& Ercolano, B.\ 2012, \aap, 539, A148 

\bibitem[Bolton et al.(2017)]{Bo17} Bolton, S.~J., Adriani, A., Adumitroaie, V., et al.\ 2017, Science, 356, 821 

\bibitem[Booth et al.(2017)]{Boo17} Booth, R.~A., Clarke, C.~J., Madhusudhan, N., \& Ilee, J.~D.\ 2017, \mnras, 469, 3994 

\bibitem[Burgdorf et al.(2003)]{Bu03} Burgdorf, M., Orton, G.~S., Davis, G.~R., et al.\ 2003, \icarus, 164, 244 

\bibitem[Ciesla(2014)]{Ci14} Ciesla, F.~J.\ 2014, \apjl, 784, L1 

\bibitem[Conrath \& Gautier(2000)]{Co00} Conrath, B.~J., \& Gautier, D.\ 2000, \icarus, 144, 124 

\bibitem[Conrath et al.(1987)]{Co87} Conrath, B., Gautier, D., Hanel, R., Lindal, G., \& Marten, A.\ 1987, \jgr, 92, 15003 

\bibitem[Dr{\c a}{\.z}kowska et al.(2016)]{Dr16} Dr{\c a}{\.z}kowska, J., Alibert, Y., \& Moore, B.\ 2016, \aap, 594, A105 

\bibitem[Fletcher et al.(2011)]{Fl11} Fletcher, L.~N., Baines, K.~H., Momary, T.~W., et al.\ 2011, \icarus, 214, 510 

\bibitem[Fletcher et al.(2009b)]{Fl09b} Fletcher, L.~N., Orton, G.~S., Teanby, N.~A., \& Irwin, P.~G.~J.\ 2009b, \icarus, 202, 543 

\bibitem[Fletcher et al.(2009a)]{Fl09a} Fletcher, L.~N., Orton, G.~S., Teanby, N.~A., Irwin, P.~G.~J., \& Bjoraker, G.~L.\ 2009a, \icarus, 199, 351

\bibitem[Gautier \& Hersant(2005)]{Ga05} Gautier, D., \& Hersant, F.\ 2005, \ssr, 116, 25 

\bibitem[Gautier et al.(2001)]{Ga01} Gautier, D., Hersant, F., Mousis, O., \& Lunine, J.~I.\ 2001, \apjl, 550, L227 

\bibitem[Gibb et al.(2004)]{Gi04} Gibb, E.~L., Whittet, D.~C.~B., Boogert, A.~C.~A., \& Tielens, A.~G.~G.~M.\ 2004, \apjs, 151, 35 

\bibitem[Guillot \& Hueso(2006)]{Gu06} Guillot, T., \& Hueso, R.\ 2006, \mnras, 367, L47 

\bibitem[Guillot \& Gladman(2000)]{Gu00} Guillot, T., \& Gladman, B.\ 2000, Disks, Planetesimals, and Planets, 219, 475 

\bibitem[Hartmann et al.(1998)]{Ha98} Hartmann, L., Calvet, N., Gullbring, E., \& D'Alessio, P.\ 1998, \apj, 495, 385 

\bibitem[Hersant et al.(2004)]{He04} Hersant, F., Gautier, D., \& Lunine, J.~I.\ 2004, \planss, 52, 623 

\bibitem[Hueso \& Guillot(2005)]{Hu05} Hueso, R., \& Guillot, T.\ 2005, \aap, 442, 703 


\bibitem[Irwin et al.(1998)]{Ir98} Irwin, P.~G.~J., Weir, A.~L., Smith, S.~E., et al.\ 1998, \jgr, 103, 23001 

\bibitem[Karkoschka \& Tomasko(2011)]{Ka11} Karkoschka, E., \& Tomasko, M.~G.\ 2011, \icarus, 211, 780 

\bibitem[Karkoschka \& Tomasko(2009)]{Ka09} Karkoschka, E., \& Tomasko, M.\ 2009, \icarus, 202, 287 

\bibitem[Kimura \& Tsuribe(2012)]{Ki12} Kimura, S.~S., \& Tsuribe, T.\ 2012, \pasj, 64, 116 

\bibitem[Kouchi(1990)]{Ko90} Kouchi, A.\ 1990, Journal of Crystal Growth, 99, 1220 

\bibitem[Lambrechts \& Johansen(2014)]{La14} Lambrechts, M., \& Johansen, A.\ 2014, \aap, 572, A107 

\bibitem[Lambrechts et al.(2014)]{Lam14} Lambrechts, M., Johansen, A., \& Morbidelli, A.\ 2014, \aap, 572, A35 

\bibitem[Lindal et al.(1990)]{Li90} Lindal, G.~F., Lyons, J.~R., Sweetnam, D.~N., et al.\ 1990, \grl, 17, 1733 

\bibitem[Lindal et al.(1987)]{Li87} Lindal, G.~F., Lyons, J.~R., Sweetnam, D.~N., et al.\ 1987, \jgr, 92, 14987 

\bibitem[Lynden-Bell \& Pringle(1974)]{Ly74} Lynden-Bell, D., \& Pringle, J.~E.\ 1974, \mnras, 168, 603 

\bibitem[Mahaffy et al.(2000)]{Ma00} Mahaffy, P.~R., Niemann, H.~B., Alpert, A., et al.\ 2000, \jgr, 105, 15061 


\bibitem[Monga \& Desch(2015)]{Mo15} Monga, N., \& Desch, S.\ 2015, \apj, 798, 9 

\bibitem[Morbidelli et al.(2016)]{Mor16} Morbidelli, A., Bitsch, B., Crida, A., et al.\ 2016, \icarus, 267, 368 

\bibitem[Mousis et al.(2018)]{Mo18} Mousis, O., Atkinson, D.~H., Cavali{\'e}, T., et al.\ 2018, \planss, 155, 12

\bibitem[Mousis et al.(2016)]{Mo16} Mousis, O., Lunine, J.~I., Luspay-Kuti, A., et al.\ 2016, \apjl, 819, L33 

\bibitem[Mousis et al.(2014)]{Mo14} Mousis, O., Lunine, J.~I., Fletcher, L.~N., et al.\ 2014, \apjl, 796, L28 

\bibitem[Mousis et al.(2012)]{Mo12} Mousis, O., Lunine, J.~I., Madhusudhan, N., \& Johnson, T.~V.\ 2012, \apjl, 751, L7 

\bibitem[Mousis et al.(2009)]{Mo09} Mousis, O., Marboeuf, U., Lunine, J.~I., et al.\ 2009, \apj, 696, 1348 

\bibitem[Nakamoto \& Nakagawa(1994)]{NN94} Nakamoto, T., \& Nakagawa, Y.\ 1994, \apj, 421, 640 

\bibitem[Niemann et al.(1998)]{Ni98} Niemann, H.~B., Atreya, S.~K., Carignan, G.~R., et al.\ 1998, \jgr, 103, 22831 

\bibitem[Orton et al.(1998)]{Or98} Orton, G.~S., Fisher, B.~M., Baines, K.~H., et al.\ 1998, \jgr, 103, 22791  

\bibitem[Owen et al.(1999)]{Ow99} Owen, T., Mahaffy, P., Niemann, H.~B., et al.\ 1999, \nat, 402, 269 

\bibitem[Pollack et al.(1996)]{Po96} Pollack, J.~B., Hubickyj, O., Bodenheimer, P., et al.\ 1996, \icarus, 124, 62 

\bibitem[Ronnet et al.(2018)]{Ro18} Ronnet, T., Mousis, O., Vernazza, P., Lunine, J.~I., \& Crida, A.\ 2018, \aj, 155, 224 


\bibitem[Shakura \& Sunyaev(1973)]{Sh73} Shakura, N.~I., \& Sunyaev, R.~A.\ 1973, \aap, 24, 337 

\bibitem[Stevenson \& Salpeter(1977b)]{St77b} Stevenson, D.~J., \& Salpeter, E.~E.\ 1977b, \apjs, 35, 239 

\bibitem[Stevenson \& Salpeter(1977a)]{St77a} Stevenson, D.~J., \& Salpeter, E.~E.\ 1977a, \apjs, 35, 221 

\bibitem[Sromovsky et al.(2014)]{Sr14} Sromovsky, L.~A., Karkoschka, E., Fry, P.~M., et al.\ 2014, \icarus, 238, 137 

\bibitem[von Zahn et al.(1998)]{vo98} von Zahn, U., Hunten, D.~M., \& Lehmacher, G.\ 1998, \jgr, 103, 22815 

\bibitem[Wilson \& Militzer(2010)]{Wi10} Wilson, H.~F., \& Militzer, B.\ 2010, Physical Review Letters, 104, 121101 

\bibitem[Wong et al.(2004)]{Wo04} Wong, M.~H., Mahaffy, P.~R., Atreya, S.~K., Niemann, H.~B., \& Owen, T.~C.\ 2004, \icarus, 171, 153 

\end{thebibliography}
\end{document}